\documentclass[twocolumn, prl, footinbib, floatfix]{revtex4-1}

\usepackage[pdftex]{graphicx}
\usepackage{amsmath,amssymb,color,subfigure,yllmath,natbib}

\begin{document}

\title{Theory of strain controlled magnetotransport and stabilization of the ferromagnetic insulating phase in manganite thin films}

\author{Anamitra Mukherjee$^{1, 2}$, William S. Cole$^2$, Patrick Woodward$^3$, Mohit Randeria$^2$, and Nandini Trivedi$^2$}

\affiliation{
$^1$Department of Physics and Astronomy, University of British Columbia, Vancouver, BC V6T 1Z1, Canada,\\
$^2$Department of Physics, The Ohio State University, Columbus, OH 43210, USA,\\
$^3$Department of Chemistry, The Ohio State University, Columbus, OH 43210, USA
}

\date{\today}
\begin{abstract}
We show that applying strain on half-doped manganites makes it possible to tune the system to the proximity of a metal-insulator transition, and thereby generate a colossal magnetoresistance (CMR) response. This phase competition not only allows control of CMR in ferromagnetic metallic manganites but can be used to generate CMR response in otherwise robust insulators at half doping. Further, from our realistic microscopic model of strain and magneto-transport calculations within the Kubo formalism, we demonstrate a striking result of strain engineering that under tensile strain a ferromagnetic charge ordered insulator,  previously inaccessible to experiments, becomes stable.
\end{abstract}
\maketitle

\emph{Introduction.}---Transition metal oxides have long been studied for their surprising emergent behavior such as high $T_c$ superconductivity in the cuprates, ferroelectricity in the titanates, and colossal magnetoresistance (CMR) in the manganites. However, very recent advances in heterojunction growth~\cite{mannhart-sci, dagotto-sci, hwang-nat-mat} have opened the possibility of producing atomically perfect interfaces of oxide materials, and therefore applying precisely controlled strain to oxide thin films. In this Letter we address the impacts of strain on ordered phases, temperature scales, and CMR in the manganites. As a specific example, we consider materials at ``half-doping" which have a prototypical chemical formula of A$_{0.5}$A'$_{0.5}$MnO$_3$ where A is a rare earth and A' an alkaline earth metal~\cite{akahoshi-prl}. 

At large bandwidths (BW), the half-doped manganites are ferromagnetic metals (F-M), while narrow BW materials are spin, charge, and orbitally ordered insulators (SCO-I) as described later in the text. CMR is known to occur in F-M materials close to the metal-insulator phase boundary, and is the result of \emph{phase competition} (between the F-M and a charge-ordered insulator), which is traditionally controlled by isovalent chemical substitutions at the A-site~\cite{akahoshi-prl}. The unstrained material has been theoretically studied extensively~\cite{dagotto-prl2, dagotto-prl-2007}.
Most prior work on the effects of strain (both theory~\cite{ahn-nat, bray-prb, dagotto-prb, strain-dft-2, millis-1, millis-2, millis-3} and
experiment~\cite{strain-mit, exp-1, wang_strong_2010, manganite-support-Tc, support-ba-bl, ps-1, ps-2, yang_enhancing_2006, konishi}) have focused on how it affects the magnetic and electronic phases, with little emphasis on magnetotransport. We extend the usual model for manganites, previously used to study magnetotransport without strain~\cite{dagotto-prl2, dagotto-prl-2007, dagotto-prb-1}, to propose and solve, for the first time, a microscopic Hamiltonian that includes the effects of strain and obtain the following results:

(i) Tensile strain provides a route to stabilizing a ferromagnetic charge ordered insulator (FC-I). This phase has not been conclusively observed in any half doped manganite with tolerance factor variations, but should finally be observable with strain engineering.

(ii) We demonstrate that strain can induce phase transitions. As a consequence, we show that the CMR in F-M materials can be enhanced by tuning the proximity to the metal-insulator transition, and that insulating phases can be made metallic under strain and therefore also exhibit a CMR response. This greatly expands the family of materials with potential device applications. 

(iii) We show that strain engineering also allows for control over $T_c$ in F-M manganites, and can be used to control the CMR temperature in the CMR materials.

\emph{Model.}---We begin with the ``standard model" for the manganites. Because of the octahedral crystal field, the Mn $e_g$ levels have a higher energy than the $t_{2g}$ levels. Combined with a large Hund's coupling, which ensures that the electron spins align ferromagnetically, this localizes three Mn $3d$ electrons in the $t_{2g}$ levels which form local $S=3/2$ moments (called ``core spins"). The remaining electrons, if any, are itinerant and occupy two bands that result from the hybridization of the Mn $e_g$ levels. The model also includes an effective antiferromagnetic superexchange $J$ between neighboring Mn
core spins, and finally the $e_g$ electrons couple to Jahn-Teller phonons with a coupling strength $\lambda$.

All energy scales are given in units of the unstrained bandwidth $t$. This Hamiltonian yields an accurate description of the physics of the manganites~\cite{hotta-review} and is discussed in detail in the Supplementary Information section I.

\begin{figure}[t]
  \centering{
  \includegraphics[width=6.5cm, height=4cm, clip=true]{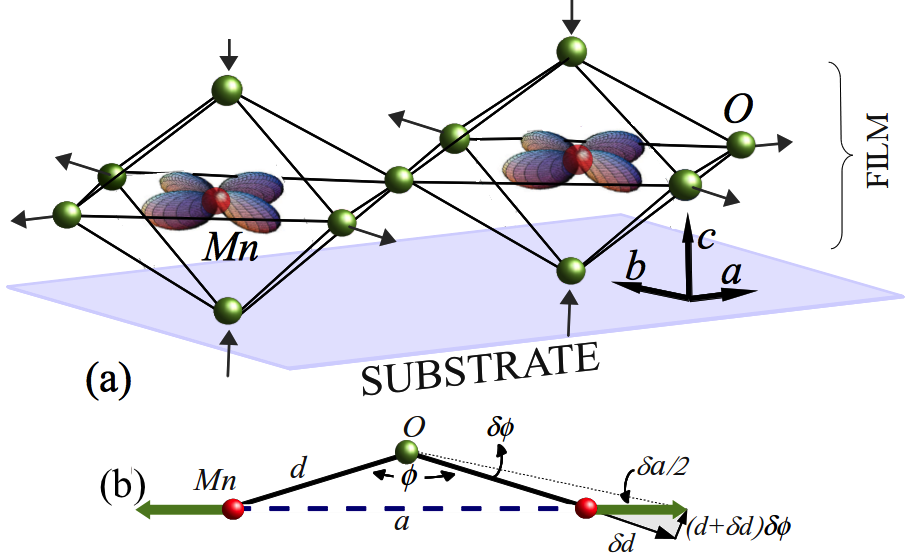}}
  \caption{
(a) Uniform tensile strain in the a-b plane is introduced by lattice mismatch with the substrate. Volume conserving tensile strain expands the in-plane Mn-O bonds while contracting bonds in the c direction. This causes higher occupancy of the in-plane $d_{x^2-y^2}$ orbital, a larger out-of-plane hopping and smaller in-plane hopping compared to the unstrained values. Compressive strain, not shown, has the opposite effect. (b) Schematic of a generic Mn-O-Mn bond under strain. We indicate the Mn-O bond length $d$, Mn-O-Mn bond angle $\phi$ and the shortest Mn-Mn distance, the lattice parameter $a$. An expansion of $a$ along the green arrows causes a change in $d$ and $\phi$.}
\vspace{-0.5cm}
  \label{f-1}
\end{figure}

We extend this model to incorporate the effects of strain. Fig.~\ref{f-1}(a) schematically illustrates substrate-induced, in-plane tensile strain. Tensile (compressive) strain is caused by growing the film on substrates  with lattice parameters larger (smaller) than those of the unstrained film. We assume that strain is applied parallel to the (a-b) plane, which we take to be the plane of the $d_{x^2-y^2}$ orbital. We quantify strain by the parameter $e_{\parallel}=[(a_s-a)/a]=\delta a/a$, where $a_s$ and $a$ are the substrate and film lattice parameters. Here $a$ is the distance between two nearest neighbor Mn atoms. We consider volume conserving strain and use the relation $e_{\bot}=-4\nu e_{\parallel}$ where $\nu$ is the Poisson ratio. We choose $\nu=0.375$, consistent with previous estimates~\cite{poisson-1, poisson-2}. In-plane compressive strain corresponds to $e_{\parallel}<0$, while $e_{\parallel}>0$ for tensile strain. With this in mind, we propose three microscopic effects of strain that must be included in our model, extending previous theoretical proposals~\cite{bray-prb, dagotto-prb}: 

(i) \emph{Strain modifies the hopping matrix elements.}
Strain affects the lattice parameter $a$ which in turn modifies both the Mn-O bond length $d$ and the Mn-O-Mn bond angle $\phi$ as seen in Fig.~\ref{f-1}(b). In the Supplementary Information section I (c), we show by using Slater-Koster~\cite{slater-koster} and Harrison~\cite{harrison}  scaling, that the effect on the hopping matrix elements due to the change in $\phi$ can be neglected for a large class of half doped manganites, and that the hopping in the (a-b) plane scales with strain as $t_{xy}  \rightarrow t_{xy}\left(1-7e_{\parallel}\right)$. We restrict our calculations to a single layer manganite film in the a-b plane (as depicted in Fig.~\ref{f-1}(a)) and refer to the unstrained in-plane hopping parameter $t_{xy}$ as $t$, and under strain as $\tilde{t}$.

(ii) \emph{Strain modifies the antiferromagnetic superexchange.}
The superexchange coupling also scales with the hopping. From similar considerations it can be shown that the in-plane superexchange scales as $J_{xy} \rightarrow (J_{\parallel}/t)(1-14e_{\parallel})$ with strain. We refer to the unstrained in-plane superexchange as $J$ and in the strained case,
$\tilde{J}$.

(iii) \emph{Strain generates an orbital bias.}
Because of the increase in the in-plane Mn-O bond length, tensile strain makes occupation of the $d_{x^2-y^2}$ orbital energetically favorable. In-plane compressive strain favors the out-of-plane $d_{3z^2-r^2}$ orbital. This orbital bias induced by in-plane compressive and tensile strains in La$_{0.7}$Sr$_{0.3}$MnO$_3$ has been observed in x-ray
absorption~\cite{orb-bias} as well as in angle resolved photoemission~\cite{orb-bias-arpes}. We incorporate this in our model
Hamiltonian with an extra term, $H_{\rm bias}=\sum_{i,\alpha}\epsilon_{\alpha}n_{i,\alpha}$
where $\epsilon_{\alpha}=\delta/2$ for $\alpha=d_{3z^2-r^2}$ and
$\epsilon_{\alpha}=-\delta/2$ for $\alpha=d_{x^2-y^2}$.

From experiments, the $e_g$ splitting has been estimated to be between $0.4t$ and $3t$~\cite{orb-bias, orb-bias-2}. We make a conservative estimate for the bias to be $\delta\approx 10e_{\parallel}t$, i.e., a splitting of $\pm0.2t$ for $\pm$2$\%$ strain. These values are consistent with density functional estimates~\cite{satpathy-strain-1, satpathy-strain-2}. Values of 2-3$\%$ for strain on manganite films, as we consider here, are easily achievable in experiments\cite{exp-1, yang_enhancing_2006}.

As mentioned, we perform our calculations in two dimensions, describing a single-layer manganite film in the a-b plane. Further, we assume strain to be uniform in the layer. This is sufficient to bring out the important features of the phase diagram and in-plane transport. We describe our method of solution in Supplementary Information sections II and III and focus here on our results. 

\begin{figure*}[t]
  \centering{
  \includegraphics[width=12cm, height=3.8cm, clip=true]{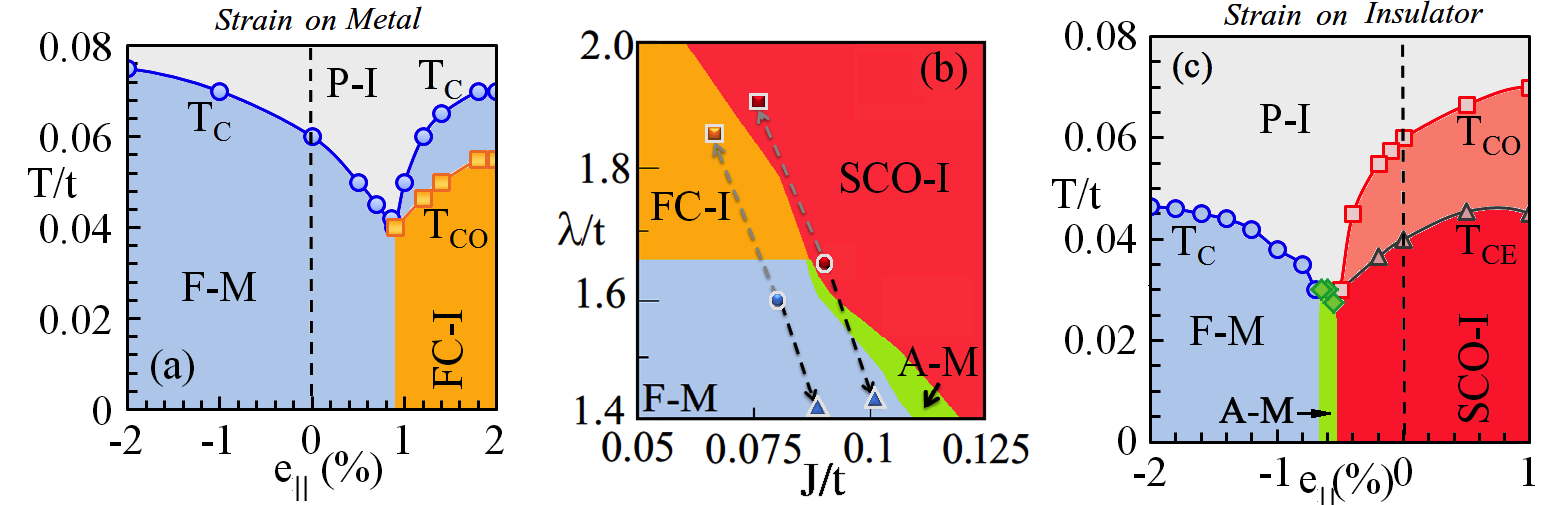}}
  \caption{
(a) The ferromagnetic metal (F-M), ferromagnetic charge ordered insulator
(FC-I), and paramagnetic insulator (P-I) in the temperature ($T$) vs.
strain ($e_{\parallel}$) plane. Compressive strain ($e_{\parallel}<0$) on F-M raises $T_c$, while tensile strain initially suppresses $T_c$ but  beyond $\sim 1\%$ strain drives the system into an FC-I. (b) Unstrained $T=0$ phase diagram in the $\lambda/t$, $J/t$ plane. This phase diagram also shows the spin-charge-orbital ordered insulator (SCO-I) and the A-type antiferromagnetic metal (A-M)~\cite{pradhan-prl}. The blue dot at ($\lambda/t$, $J/t$)=(1.6, 0.08) refers to the unstrained starting point from which we calculate (a), while the red dot at ($\lambda/t$, $J/t$)=(1.65, 0.09) refers to the unstrained point from which we calculate the phase diagram (c). Under strain, the hopping parameter $t$ is rescaled which changes both the ratios $\lambda/t$ and $J/t$. The symbols at the arrow heads indicate such values for 2$\%$ compressive (triangles) and tensile (squares) strain respectively. (c) Tensile strain on the SCO-I enhances charge ordering temperature $T_{CO}$ while compressive strain causes a transition to F-M. $T_{CE}$ denotes the spin ordering temperature. In (a) and (c), for clarity, we do not show strain-induced low temperature equilibrium phase separation~\cite{ps-status}.}
\vspace{-0.5cm}
\label{f-2}
\end{figure*}

\emph{Strain driven phase transitions.}---Fig.~\ref{f-2}(b) shows the $T=0$ $\lambda-J$ phase diagram without strain. On this we denote two representative parameter points, F-M (blue dot) and SCO-I (red dot). The SCO-I is an insulator with planar checkerboard charge order (CO), alternating $d_{x^2-r^2}/d_{y^2-r^2}$ orbital order (OO) (on the sites with larger charge density), and CE type spin order (zig-zag ferromagnetic chains coupled antiferromagnetically). Fig.~\ref{f-2}(a) and (c) show the effect of strain on these points.

Starting from these parameters, in-plane compressive strain favors F-M as
seen in both (a) and (c). There are two competing effects here, however.
Compressive strain increases the in-plane hopping which in turn reduces
$\lambda/\tilde{t}$, as seen by following the dashed  black arrow in (b). This favors a metallic state where double-exchange promotes ferromagnetism. On the other hand, compressive strain increases $\tilde{J}/\tilde{t}$, which tends to narrow the BW, while the orbital bias promotes occupancy of the out-of-plane $d_{3z^2-r^2}$ orbital. Both of these latter effects work against the stability of the F-M, but we find the F-M to be dominant up to the maximum values of strain we have considered.

In-plane tensile strain stabilizes insulators that can have either
ferromagnetic or antiferromagnetic spin textures as seen in (a) and (c)
respectively. The insulators have long-range checkerboard charge order and are stabilized by the reduced in-plane hybridization or increased $\lambda/\tilde{t}$, as seen by following the grey dashed arrow in (b). This effect tends to localize the electrons. While sufficient increase in $\lambda$ eventually turns the system insulating regardless of the unstrained F-M parameter, the magnetic order depends crucially on the value of $\tilde{J}/\tilde{t}$.

This dependence of the magnetic/charge-ordering scales on strain has been seen in experiments, both away from~\cite{manganite-support-Tc} and at half-doping~\cite{yang_enhancing_2006, exp-1}. They include increasing $T_c$ with compressive strain in a $La_{0.8}$Ba$_{0.2}$MnO$_3$, suppressing $T_c$ with tensile strain in La$_{0.67}$Ca$_{0.33}$MnO$_3$, and increasing $T_{CO}$ with tensile strain on Pr$_{0.5}$Ca$_{0.5}$MnO$_3$~\cite{yang_enhancing_2006, exp-1}. 
We note that while there is a dearth of experimental data on scaling of t and J for manganites, our results are robust to typical variations in the scaling~\cite{scaling-status}.

\emph{Stability of the FC-I phase.}---From Fig.~\ref{f-2} (b) we see that
adequate tensile strain on F-M with J/t$\sim$0.05-0.08 convert
the system into a FC-I, just as that depicted for ($\lambda/t=1.6$, $J/t=0.08$) by the grey dashed arrow. The unstrained FC-I phase, was predicted in theory~\cite{yunoki} at $\lambda, J$ values as in Fig.~\ref{f-2} (b). In small BW half-doped manganites, \textit{e.g.} La$_{0.5}$Ca$_{0.5}$MnO$_3$, signatures of this phase coexisting with AF-CO phase were reported~\cite{mathur} at 90K. This implies that in the half-doped manganites either FC-I is the true ground state only in a narrow $\lambda, J$ widow or it is a metastable state with energy very close to the true ground state. We predict that tensile strain on an ordered F-M suppresses other phases and can stabilize the FC-I as the ground state.

\emph{Effect of strain on magnetotransport.}---The maximum CMR temperature achievable by BW tuned phase competition is the $T_C$ of the unstrained material. Additionally, bicritical nature of the phase diagram and proximity to the metal-insulator boundary needed for CMR, keeps the $T_C$ quite low~\cite{tok-book}. We show that because strain affects different intrinsic energy scales differently, it not only tunes phase competition, but also allows optimization of the competition between CMR temperature and $\%$MR.

\begin{figure}[t]
  \centering
  \includegraphics[width=8cm, height=6.0cm, clip=true]{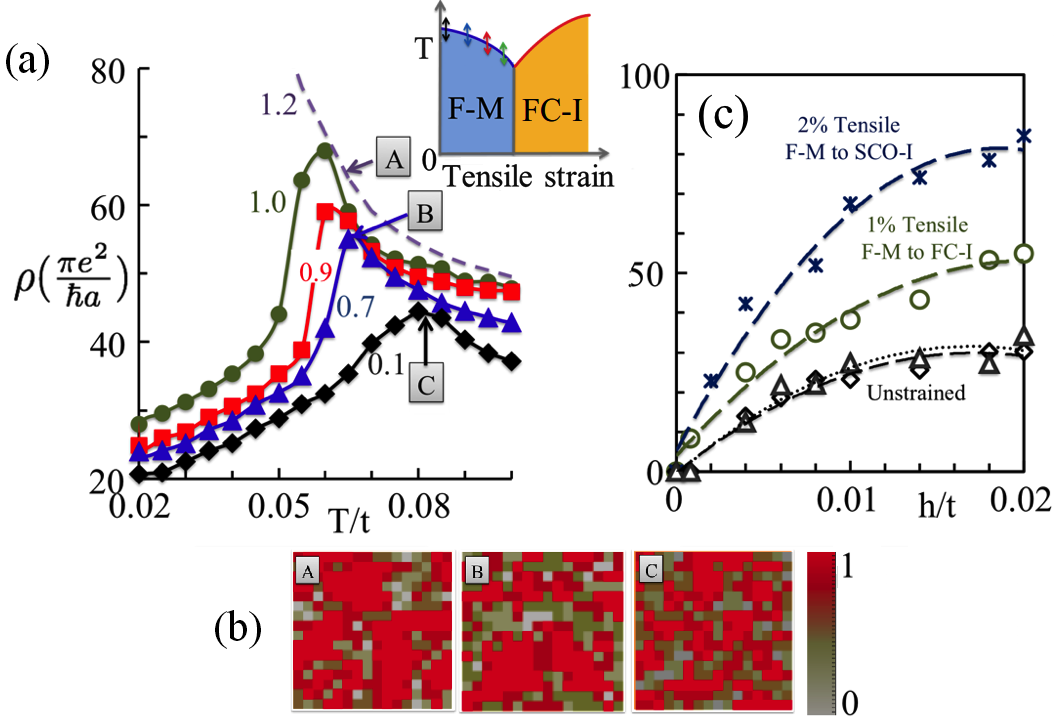}
  \caption{(a) Resistivity $\rho(T)$ for different values of tensile strain. Strain values are marked on the individual curves, which are color coded with the arrows in the inset (which is a schematic of
Fig.~\ref{f-2}(a) showing only the tensile strain part). Note that while the
temperature at which the peak occurs (which is close to $T_C$) decreases, the peak value of the resistivity increases. (b) Real space snapshots of  charge ordering at $T_C$; the volume fraction of checkerboard charge ordered regions (shown in red) grows systematically with tensile strain. The labels A, B and C on them and in (a) denote the strain, temperature values where these were calculated. The magnetoresistance ($\%MR$) as a function of magnetic field for $\sim$1$\%$ strain on a system close to the F-M/FC-I phase boundary and $\sim$2$\%$ on a system close to the F-M/SCO-I boundary. The unstrained values for both parent parameter points (triangles and diamonds) are shown for comparison.}
\vspace{-0.5cm}
  \label{f-3}
\end{figure}

Fig.~\ref{f-3}(a) shows the resistivity, $\rho(T)$, for various tensile strain values on the unstrained F-M phase (blue dot in Fig.~\ref{f-2}(b)). While  CMR behavior has been reported before~\cite{dagotto-prl2, dagotto-prl-2007, dagotto-prb}, our novelty is the use of strain as an external knob. Increasing tensile strain causes rapid rise in the resistivity maximum that occurs at $T\sim T_C$, accompanied with reduction of both the $T_C$ and the temperature at the resistivity maximum ($T_{CMR}$). The reduction in $T_C$ is due to the approach to the F-M/SCO-I boundary by increasing the tensile strain, as depicted in the inset in (a). The reason for the increase in the resistivity maximum is the strain-induced enhancement of metal-insulator coexistence at $T_C$ as illustrated in (b). 

The color maps here depict the volume fraction of the insulating regions (red patches) embedded in an otherwise conducting background at $T_C$. These insulating regions grow in volume with increasing strain and have short range ($\pi,\pi$) CO correlations; the same correlations that one finds in the competing FC-I phase. The thermal fluctuations at $T_C$ are typically dominated by the nearest metastable minimum, in this case the FC-I phase.  
Further, since strain controls the proximity to the F-M/FC-I boundary,
increasing tensile strain makes the FC-I state progressively approach the
energy of the F-M ground state, favoring greater insulating regions with short range ($\pi,\pi$) CO correlations. 
If we start with other initial (unstrained) starting points, tensile strain can result in a F-M to SCO-I phase transition. The qualitative behavior of  magnetotransport is the same as in the F-M to FC-I case shown here. Magnetotransport data near the FM/SCO-I phase boundary is shown in  Supplementary Information Section IV.

In Fig.~\ref{f-3}(c) we show the $\%MR$, defined as $100 \times
[\rho(0)-\rho(h)]/\rho(0)$ and calculated at $T=T_{CMR}$, as a function of
magnetic field for two cases. One shows $\%$MR close to the
F-M/FC-I phase boundary, with (circles) and without (diamonds) strain; the other shows $\%$MR close to the F-M/SCO-I boundary. The amount of increase in the resistivity maximum with strain and the $\%MR$ depends on the domains of metastability of the competing phases, the type of the insulator and the largest tensile strain that can be applied before the system becomes insulating. However regardless of the nature of metal-insulator phase competition, applying tensile strain yields an enormous enhancement of $\%$MR (circles and stars) over the unstrained values (diamonds and triangles).

\begin{figure}[t]
  \centering
  \includegraphics[width=8cm, height=4cm, clip=true]{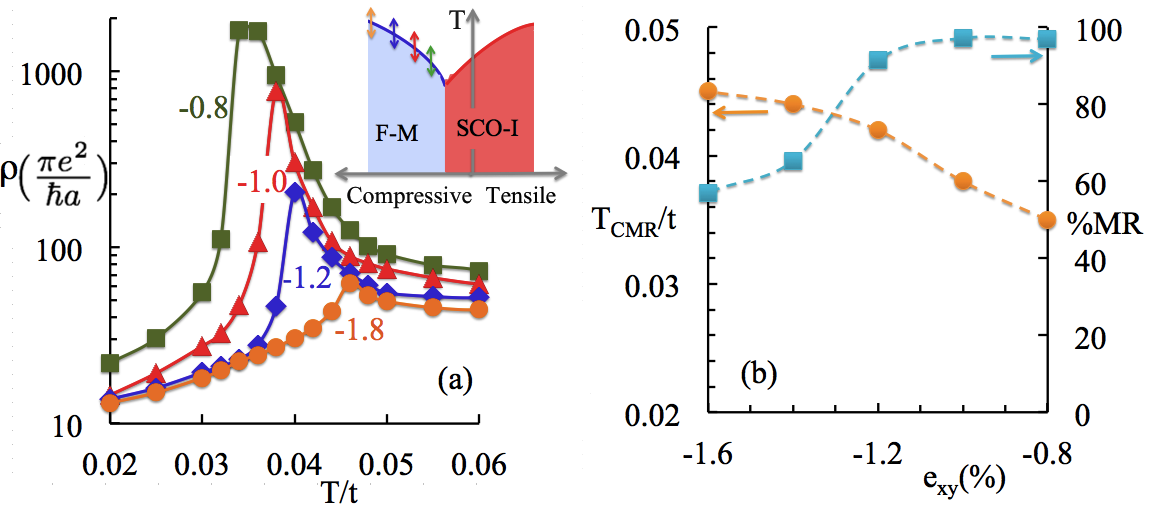}
  \caption{
Compressive strain can induce F-M in insulating manganites: (a) shows $\rho(T)$ of the resulting F-M as a function of temperature for various values of compression (color coded with the double arrows in the inset, which is a schematic of Fig.~\ref{f-2}(c)). There is a pronounced peak which systematically shifts to higher temperatures with increasing compression. (b) Temperature at resistivity peak $T_{CMR}$ (left axis), which increases with increasing compression, and $\%$MR (right axis),  which decreases as compression is increased. The $\%$MR does not decrease drastically at first, and there is a region where $T_{CMR}$ can be enhanced without losing the large $\%$MR.}
\vspace{-0.5cm}
  \label{f-4}
\end{figure}

Finally we demonstrate that \emph{compressive} strain can drive insulators
across the metal-insulator transition into the CMR regime. Fig.~\ref{f-4}(a)
shows $\rho(T)$ with increasing compressive strain on SCO-I. At 0.8$\%$ strain, the insulator-to-metal transition is accompanied by a CE-to-ferromagnetic transition. Increasing the compressive strain further causes a monotonic increase in $T_C$ (also seen in the inset in (a)). The peak in the resistivity with increasing strain is also systematically shifted to higher temperatures. In (b) we plot $\%$MR at the temperature of the resistivity maximum as a function of strain. We also show the corresponding CMR temperatures ($T_{CMR}$).  We find that $\%MR$ is reduced upon increasing strain, as expected, but there is in fact an optimal region in which $T_{CMR}$ can be increased without substantially reducing $\%MR$. We have checked that our results survive A site disordering.

\emph{Conclusion.}---We stress a major difference between the strain engineering  and isovalent substitution. While both can tune the bandwidth, uniform strain will not introduce the short-range disorder that naturally results from substitution. As a result, ferromagnetic $T_C$'s and $T_{CMR}$ of strained films should be higher than in their substitution-engineered counterparts. In turn, larger intrinsic bandwidth would cause greater $\%$MR at smaller magnetic fields, since the external field required to align the core spins is reduced. Also, strain effects on manganites of any material composition and doping can be directly studied in our approach. 
Our results indicate some promising future directions. First, we have
demonstrated that strain gives access to a large phase space of new and
accessible states for a given unstrained material. Second, strain need not introduce disorder, in contrast to chemical substitution. Finally, strain directly impacts orbital occupancy in a tunable way, and opens new possibilities for orbital-state sensitive electronics.

We gratefully acknowledge support from
DOE DE-FG02-07ER46423 (A.M.),
DOE BES DE-SC0005035 (M.R.),
NSF DMR-0907275 (N.T.), and Center for Emergent Materials, 
NSF MRSEC DMR-0820414 (W.S.C. and P.W.).

\bibliography{mnbib7.bib}

\begin{thebibliography}{40}%
\makeatletter
\providecommand \@ifxundefined [1]{%
 \@ifx{#1\undefined}
}%
\providecommand \@ifnum [1]{%
 \ifnum #1\expandafter \@firstoftwo
 \else \expandafter \@secondoftwo
 \fi
}%
\providecommand \@ifx [1]{%
 \ifx #1\expandafter \@firstoftwo
 \else \expandafter \@secondoftwo
 \fi
}%
\providecommand \natexlab [1]{#1}%
\providecommand \enquote  [1]{``#1''}%
\providecommand \bibnamefont  [1]{#1}%
\providecommand \bibfnamefont [1]{#1}%
\providecommand \citenamefont [1]{#1}%
\providecommand \href@noop [0]{\@secondoftwo}%
\providecommand \href [0]{\begingroup \@sanitize@url \@href}%
\providecommand \@href[1]{\@@startlink{#1}\@@href}%
\providecommand \@@href[1]{\endgroup#1\@@endlink}%
\providecommand \@sanitize@url [0]{\catcode `\\12\catcode `\$12\catcode
  `\&12\catcode `\#12\catcode `\^12\catcode `\_12\catcode `\%12\relax}%
\providecommand \@@startlink[1]{}%
\providecommand \@@endlink[0]{}%
\providecommand \url  [0]{\begingroup\@sanitize@url \@url }%
\providecommand \@url [1]{\endgroup\@href {#1}{\urlprefix }}%
\providecommand \urlprefix  [0]{URL }%
\providecommand \Eprint [0]{\href }%
\providecommand \doibase [0]{http://dx.doi.org/}%
\providecommand \selectlanguage [0]{\@gobble}%
\providecommand \bibinfo  [0]{\@secondoftwo}%
\providecommand \bibfield  [0]{\@secondoftwo}%
\providecommand \translation [1]{[#1]}%
\providecommand \BibitemOpen [0]{}%
\providecommand \bibitemStop [0]{}%
\providecommand \bibitemNoStop [0]{.\EOS\space}%
\providecommand \EOS [0]{\spacefactor3000\relax}%
\providecommand \BibitemShut  [1]{\csname bibitem#1\endcsname}%
\let\auto@bib@innerbib\@empty
\bibitem [{\citenamefont {Mannhart}\ \emph {et~al.}(2010)\citenamefont
  {Mannhart} \emph {et~al.}}]{mannhart-sci}%
  \BibitemOpen
  \bibfield  {author} {\bibinfo {author} {\bibfnamefont {J.}~\bibnamefont
  {Mannhart}} \emph {et~al.},\ }\href@noop {} {\bibfield  {journal} {\bibinfo
  {journal} {Science}\ }\textbf {\bibinfo {volume} {327}},\ \bibinfo {pages}
  {1607} (\bibinfo {year} {2010})}\BibitemShut {NoStop}%
\bibitem [{\citenamefont {Dagotto}(2007)}]{dagotto-sci}%
  \BibitemOpen
  \bibfield  {author} {\bibinfo {author} {\bibfnamefont {E.}~\bibnamefont
  {Dagotto}},\ }\href@noop {} {\bibfield  {journal} {\bibinfo  {journal}
  {Science}\ }\textbf {\bibinfo {volume} {318}},\ \bibinfo {pages} {1076}
  (\bibinfo {year} {2007})}\BibitemShut {NoStop}%
\bibitem [{\citenamefont {Hwang}\ \emph {et~al.}(2012)\citenamefont {Hwang}
  \emph {et~al.}}]{hwang-nat-mat}%
  \BibitemOpen
  \bibfield  {author} {\bibinfo {author} {\bibfnamefont {H.~Y.}\ \bibnamefont
  {Hwang}} \emph {et~al.},\ }\href@noop {} {\bibfield  {journal} {\bibinfo
  {journal} {Nature Materials}\ }\textbf {\bibinfo {volume} {11}},\ \bibinfo
  {pages} {103} (\bibinfo {year} {2012})}\BibitemShut {NoStop}%
\bibitem [{\citenamefont {Akahoshi}\ \emph {et~al.}(2003)\citenamefont
  {Akahoshi} \emph {et~al.}}]{akahoshi-prl}%
  \BibitemOpen
  \bibfield  {author} {\bibinfo {author} {\bibfnamefont {D.}~\bibnamefont
  {Akahoshi}} \emph {et~al.},\ }\href@noop {} {\bibfield  {journal} {\bibinfo
  {journal} {Phys. Rev. Lett.}\ }\textbf {\bibinfo {volume} {90}},\ \bibinfo
  {pages} {177203} (\bibinfo {year} {2003})}\BibitemShut {NoStop}%
\bibitem [{\citenamefont {\ifmmode~\mbox{\c{S}}\else \c{S}\fi{}en}\ \emph
  {et~al.}(2010)\citenamefont {\ifmmode~\mbox{\c{S}}\else \c{S}\fi{}en} \emph
  {et~al.}}]{dagotto-prl2}%
  \BibitemOpen
  \bibfield  {author} {\bibinfo {author} {\bibfnamefont {C.}~\bibnamefont
  {\ifmmode~\mbox{\c{S}}\else \c{S}\fi{}en}} \emph {et~al.},\ }\href@noop {}
  {\bibfield  {journal} {\bibinfo  {journal} {Phys. Rev. Lett.}\ }\textbf
  {\bibinfo {volume} {105}},\ \bibinfo {pages} {097203} (\bibinfo {year}
  {2010})}\BibitemShut {NoStop}%
\bibitem [{\citenamefont {\ifmmode~\mbox{\c{S}}\else \c{S}\fi{}en}\ \emph
  {et~al.}(2007)\citenamefont {\ifmmode~\mbox{\c{S}}\else \c{S}\fi{}en} \emph
  {et~al.}}]{dagotto-prl-2007}%
  \BibitemOpen
  \bibfield  {author} {\bibinfo {author} {\bibfnamefont {C.}~\bibnamefont
  {\ifmmode~\mbox{\c{S}}\else \c{S}\fi{}en}} \emph {et~al.},\ }\href@noop {}
  {\bibfield  {journal} {\bibinfo  {journal} {Phys. Rev. Lett.}\ }\textbf
  {\bibinfo {volume} {98}},\ \bibinfo {pages} {127202} (\bibinfo {year}
  {2007})}\BibitemShut {NoStop}%
\bibitem [{\citenamefont {Ahn}\ \emph {et~al.}(2004)\citenamefont {Ahn} \emph
  {et~al.}}]{ahn-nat}%
  \BibitemOpen
  \bibfield  {author} {\bibinfo {author} {\bibfnamefont {K.~H.}\ \bibnamefont
  {Ahn}} \emph {et~al.},\ }\href@noop {} {\bibfield  {journal} {\bibinfo
  {journal} {Nature}\ }\textbf {\bibinfo {volume} {428}},\ \bibinfo {pages}
  {401} (\bibinfo {year} {2004})}\BibitemShut {NoStop}%
\bibitem [{\citenamefont {Baena}\ \emph {et~al.}(2011)\citenamefont {Baena}
  \emph {et~al.}}]{bray-prb}%
  \BibitemOpen
  \bibfield  {author} {\bibinfo {author} {\bibfnamefont {A.}~\bibnamefont
  {Baena}} \emph {et~al.},\ }\href@noop {} {\bibfield  {journal} {\bibinfo
  {journal} {Phys. Rev. B}\ }\textbf {\bibinfo {volume} {83}},\ \bibinfo
  {pages} {064424} (\bibinfo {year} {2011})}\BibitemShut {NoStop}%
\bibitem [{\citenamefont {Dong}\ \emph {et~al.}(2010)\citenamefont {Dong} \emph
  {et~al.}}]{dagotto-prb}%
  \BibitemOpen
  \bibfield  {author} {\bibinfo {author} {\bibfnamefont {S.}~\bibnamefont
  {Dong}} \emph {et~al.},\ }\href@noop {} {\bibfield  {journal} {\bibinfo
  {journal} {Phys. Rev. B}\ }\textbf {\bibinfo {volume} {82}},\ \bibinfo
  {pages} {035118} (\bibinfo {year} {2010})}\BibitemShut {NoStop}%
\bibitem [{\citenamefont {Lee}\ \emph {et~al.}(2010)\citenamefont {Lee} \emph
  {et~al.}}]{strain-dft-2}%
  \BibitemOpen
  \bibfield  {author} {\bibinfo {author} {\bibfnamefont {J.~H.}\ \bibnamefont
  {Lee}} \emph {et~al.},\ }\href@noop {} {\bibfield  {journal} {\bibinfo
  {journal} {Phys. Rev. Lett.}\ }\textbf {\bibinfo {volume} {104}},\ \bibinfo
  {pages} {207204} (\bibinfo {year} {2010})}\BibitemShut {NoStop}%
\bibitem [{\citenamefont {Ahn}\ \emph {et~al.}(2001)\citenamefont {Ahn} \emph
  {et~al.}}]{millis-1}%
  \BibitemOpen
  \bibfield  {author} {\bibinfo {author} {\bibfnamefont {K.~H.}\ \bibnamefont
  {Ahn}} \emph {et~al.},\ }\href@noop {} {\bibfield  {journal} {\bibinfo
  {journal} {Physical Review B}\ }\textbf {\bibinfo {volume} {64}},\ \bibinfo
  {pages} {115103} (\bibinfo {year} {2001})}\BibitemShut {NoStop}%
\bibitem [{\citenamefont {Millis}\ \emph {et~al.}(1998)\citenamefont {Millis}
  \emph {et~al.}}]{millis-2}%
  \BibitemOpen
  \bibfield  {author} {\bibinfo {author} {\bibfnamefont {A.~J.}\ \bibnamefont
  {Millis}} \emph {et~al.},\ }\href@noop {} {\bibfield  {journal} {\bibinfo
  {journal} {Journal of Applied Physics}\ }\textbf {\bibinfo {volume} {83}},\
  \bibinfo {pages} {1588} (\bibinfo {year} {1998})}\BibitemShut {NoStop}%
\bibitem [{\citenamefont {Calderón}\ \emph {et~al.}(2003)\citenamefont
  {Calderón} \emph {et~al.}}]{millis-3}%
  \BibitemOpen
  \bibfield  {author} {\bibinfo {author} {\bibfnamefont {M.~J.}\ \bibnamefont
  {Calderón}} \emph {et~al.},\ }\href@noop {} {\bibfield  {journal} {\bibinfo
  {journal} {Physical Review B}\ }\textbf {\bibinfo {volume} {68}},\ \bibinfo
  {pages} {100401} (\bibinfo {year} {2003})}\BibitemShut {NoStop}%
\bibitem [{\citenamefont {Yang}\ \emph {et~al.}(2010)\citenamefont {Yang} \emph
  {et~al.}}]{strain-mit}%
  \BibitemOpen
  \bibfield  {author} {\bibinfo {author} {\bibfnamefont {F.}~\bibnamefont
  {Yang}} \emph {et~al.},\ }\href@noop {} {\bibfield  {journal} {\bibinfo
  {journal} {Appl. Phys. Lett.}\ }\textbf {\bibinfo {volume} {97}},\ \bibinfo
  {pages} {092503} (\bibinfo {year} {2010})}\BibitemShut {NoStop}%
\bibitem [{\citenamefont {Okuyama}\ \emph {et~al.}(2009)\citenamefont {Okuyama}
  \emph {et~al.}}]{exp-1}%
  \BibitemOpen
  \bibfield  {author} {\bibinfo {author} {\bibfnamefont {D.}~\bibnamefont
  {Okuyama}} \emph {et~al.},\ }\href@noop {} {\bibfield  {journal} {\bibinfo
  {journal} {Appl. Phys. Lett.}\ }\textbf {\bibinfo {volume} {95}},\ \bibinfo
  {pages} {152502} (\bibinfo {year} {2009})}\BibitemShut {NoStop}%
\bibitem [{\citenamefont {Wang}\ \emph {et~al.}(2010)\citenamefont {Wang} \emph
  {et~al.}}]{wang_strong_2010}%
  \BibitemOpen
  \bibfield  {author} {\bibinfo {author} {\bibfnamefont {J.}~\bibnamefont
  {Wang}} \emph {et~al.},\ }\href@noop {} {\bibfield  {journal} {\bibinfo
  {journal} {Appl. Phys. Lett.}\ }\textbf {\bibinfo {volume} {96}},\ \bibinfo
  {pages} {052501} (\bibinfo {year} {2010})}\BibitemShut {NoStop}%
\bibitem [{\citenamefont {Chou}\ \emph {et~al.}(2006)\citenamefont {Chou} \emph
  {et~al.}}]{manganite-support-Tc}%
  \BibitemOpen
  \bibfield  {author} {\bibinfo {author} {\bibfnamefont {H.}~\bibnamefont
  {Chou}} \emph {et~al.},\ }\href@noop {} {\bibfield  {journal} {\bibinfo
  {journal} {Appl. Phys. Lett.}\ }\textbf {\bibinfo {volume} {89}},\ \bibinfo
  {pages} {082511} (\bibinfo {year} {2006})}\BibitemShut {NoStop}%
\bibitem [{\citenamefont {Xie}\ \emph {et~al.}(2008)\citenamefont {Xie} \emph
  {et~al.}}]{support-ba-bl}%
  \BibitemOpen
  \bibfield  {author} {\bibinfo {author} {\bibfnamefont {C.~K.}\ \bibnamefont
  {Xie}} \emph {et~al.},\ }\href@noop {} {\bibfield  {journal} {\bibinfo
  {journal} {Appl. Phys. Lett.}\ }\textbf {\bibinfo {volume} {93}},\ \bibinfo
  {pages} {182507} (\bibinfo {year} {2008})}\BibitemShut {NoStop}%
\bibitem [{\citenamefont {Pena}\ \emph {et~al.}(2006)\citenamefont {Pena},
  \citenamefont {Sefrioui}, \citenamefont {Arias}, \citenamefont {Leon},
  \citenamefont {Santamaria}, \citenamefont {Varela}, \citenamefont
  {Pennycook}, \citenamefont {Garcia-Hernandez},\ and\ \citenamefont
  {Martinez}}]{ps-1}%
  \BibitemOpen
  \bibfield  {author} {\bibinfo {author} {\bibfnamefont {V.}~\bibnamefont
  {Pena}}, \bibinfo {author} {\bibfnamefont {Z.}~\bibnamefont {Sefrioui}},
  \bibinfo {author} {\bibfnamefont {D.}~\bibnamefont {Arias}}, \bibinfo
  {author} {\bibfnamefont {C.}~\bibnamefont {Leon}}, \bibinfo {author}
  {\bibfnamefont {J.}~\bibnamefont {Santamaria}}, \bibinfo {author}
  {\bibfnamefont {M.}~\bibnamefont {Varela}}, \bibinfo {author} {\bibfnamefont
  {S.}~\bibnamefont {Pennycook}}, \bibinfo {author} {\bibfnamefont
  {M.}~\bibnamefont {Garcia-Hernandez}}, \ and\ \bibinfo {author}
  {\bibfnamefont {J.}~\bibnamefont {Martinez}},\ }\href {\doibase
  10.1016/j.jpcs.2005.10.022} {\bibfield  {journal} {\bibinfo  {journal}
  {Journal of Physics and Chemistry of Solids}\ }\textbf {\bibinfo {volume}
  {67}},\ \bibinfo {pages} {472 } (\bibinfo {year} {2006})}\BibitemShut
  {NoStop}%
\bibitem [{\citenamefont {de~Brion}\ \emph {et~al.}(2004)\citenamefont
  {de~Brion}, \citenamefont {Chouteau}, \citenamefont {Janossy}, \citenamefont
  {Buzin},\ and\ \citenamefont {Prellier}}]{ps-2}%
  \BibitemOpen
  \bibfield  {author} {\bibinfo {author} {\bibfnamefont {S.}~\bibnamefont
  {de~Brion}}, \bibinfo {author} {\bibfnamefont {G.}~\bibnamefont {Chouteau}},
  \bibinfo {author} {\bibfnamefont {A.}~\bibnamefont {Janossy}}, \bibinfo
  {author} {\bibfnamefont {E.~R.}\ \bibnamefont {Buzin}}, \ and\ \bibinfo
  {author} {\bibfnamefont {W.}~\bibnamefont {Prellier}},\ }\href {\doibase
  10.1016/j.jmmm.2003.11.155} {\bibfield  {journal} {\bibinfo  {journal}
  {Journal of Magnetism and Magnetic Materials}\ }\textbf {\bibinfo {volume}
  {272 - 276, Part 1}},\ \bibinfo {pages} {450 } (\bibinfo {year}
  {2004})}\BibitemShut {NoStop}%
\bibitem [{\citenamefont {Yang}\ \emph {et~al.}(2006)\citenamefont {Yang} \emph
  {et~al.}}]{yang_enhancing_2006}%
  \BibitemOpen
  \bibfield  {author} {\bibinfo {author} {\bibfnamefont {Z.~Q.}\ \bibnamefont
  {Yang}} \emph {et~al.},\ }\href@noop {} {\bibfield  {journal} {\bibinfo
  {journal} {Applied Physics Letters}\ }\textbf {\bibinfo {volume} {88}},\
  \bibinfo {pages} {072507} (\bibinfo {year} {2006})}\BibitemShut {NoStop}%
\bibitem [{\citenamefont {Konishi}\ \emph {et~al.}(1999)\citenamefont {Konishi}
  \emph {et~al.}}]{konishi}%
  \BibitemOpen
  \bibfield  {author} {\bibinfo {author} {\bibfnamefont {Y.}~\bibnamefont
  {Konishi}} \emph {et~al.},\ }\href@noop {} {\bibfield  {journal} {\bibinfo
  {journal} {J. Phys. Soc. Jpn.}\ }\textbf {\bibinfo {volume} {68}},\ \bibinfo
  {pages} {3790} (\bibinfo {year} {1999})}\BibitemShut {NoStop}%
\bibitem [{\citenamefont {Yu}\ \emph {et~al.}(2008)\citenamefont {Yu} \emph
  {et~al.}}]{dagotto-prb-1}%
  \BibitemOpen
  \bibfield  {author} {\bibinfo {author} {\bibfnamefont {R.}~\bibnamefont {Yu}}
  \emph {et~al.},\ }\href {\doibase 10.1103/PhysRevB.77.214434} {\bibfield
  {journal} {\bibinfo  {journal} {Phys. Rev. B}\ }\textbf {\bibinfo {volume}
  {77}},\ \bibinfo {pages} {214434} (\bibinfo {year} {2008})}\BibitemShut
  {NoStop}%
\bibitem [{\citenamefont {Dagotto}\ \emph {et~al.}(2001)\citenamefont {Dagotto}
  \emph {et~al.}}]{hotta-review}%
  \BibitemOpen
  \bibfield  {author} {\bibinfo {author} {\bibfnamefont {E.}~\bibnamefont
  {Dagotto}} \emph {et~al.},\ }\href@noop {} {\bibfield  {journal} {\bibinfo
  {journal} {Physics Reports}\ }\textbf {\bibinfo {volume} {344}},\ \bibinfo
  {pages} {1} (\bibinfo {year} {2001})}\BibitemShut {NoStop}%
\bibitem [{\citenamefont {Yamada}\ \emph {et~al.}(2006)\citenamefont {Yamada}
  \emph {et~al.}}]{poisson-1}%
  \BibitemOpen
  \bibfield  {author} {\bibinfo {author} {\bibfnamefont {H.}~\bibnamefont
  {Yamada}} \emph {et~al.},\ }\href@noop {} {\bibfield  {journal} {\bibinfo
  {journal} {Appl. Phys. Lett.}\ }\textbf {\bibinfo {volume} {89}},\ \bibinfo
  {pages} {052506} (\bibinfo {year} {2006})}\BibitemShut {NoStop}%
\bibitem [{\citenamefont {Adamo}\ \emph {et~al.}(2009)\citenamefont {Adamo}
  \emph {et~al.}}]{poisson-2}%
  \BibitemOpen
  \bibfield  {author} {\bibinfo {author} {\bibfnamefont {C.}~\bibnamefont
  {Adamo}} \emph {et~al.},\ }\href@noop {} {\bibfield  {journal} {\bibinfo
  {journal} {Appl. Phys. Lett.}\ }\textbf {\bibinfo {volume} {95}},\ \bibinfo
  {pages} {112504} (\bibinfo {year} {2009})}\BibitemShut {NoStop}%
\bibitem [{\citenamefont {Slater}\ and\ \citenamefont
  {Koster}(1954)}]{slater-koster}%
  \BibitemOpen
  \bibfield  {author} {\bibinfo {author} {\bibfnamefont {J.~C.}\ \bibnamefont
  {Slater}}\ and\ \bibinfo {author} {\bibfnamefont {G.~F.}\ \bibnamefont
  {Koster}},\ }\href@noop {} {\bibfield  {journal} {\bibinfo  {journal} {Phys.
  Rev.}\ }\textbf {\bibinfo {volume} {94}},\ \bibinfo {pages} {1498} (\bibinfo
  {year} {1954})}\BibitemShut {NoStop}%
\bibitem [{\citenamefont {Harrison}(1989)}]{harrison}%
  \BibitemOpen
  \bibfield  {author} {\bibinfo {author} {\bibfnamefont {W.}~\bibnamefont
  {Harrison}},\ }\href@noop {} {\bibfield  {journal} {\bibinfo  {journal}
  {\textit{Electronic Structure and the Properties of Solids: The Physics of
  the Chemical Bond, Dover}}\ } (\bibinfo {year} {1989})}\BibitemShut {NoStop}%
\bibitem [{\citenamefont {Aruta}\ \emph {et~al.}(2006)\citenamefont {Aruta}
  \emph {et~al.}}]{orb-bias}%
  \BibitemOpen
  \bibfield  {author} {\bibinfo {author} {\bibfnamefont {C.}~\bibnamefont
  {Aruta}} \emph {et~al.},\ }\href@noop {} {\bibfield  {journal} {\bibinfo
  {journal} {Phys. Rev. B}\ }\textbf {\bibinfo {volume} {73}},\ \bibinfo
  {pages} {235121} (\bibinfo {year} {2006})}\BibitemShut {NoStop}%
\bibitem [{\citenamefont {Tebano}\ \emph {et~al.}(2010)\citenamefont {Tebano}
  \emph {et~al.}}]{orb-bias-arpes}%
  \BibitemOpen
  \bibfield  {author} {\bibinfo {author} {\bibfnamefont {A.}~\bibnamefont
  {Tebano}} \emph {et~al.},\ }\href@noop {} {\bibfield  {journal} {\bibinfo
  {journal} {Phys. Rev. B}\ }\textbf {\bibinfo {volume} {82}},\ \bibinfo
  {pages} {214407} (\bibinfo {year} {2010})}\BibitemShut {NoStop}%
\bibitem [{\citenamefont {Tebano}\ \emph {et~al.}(2006)\citenamefont {Tebano}
  \emph {et~al.}}]{orb-bias-2}%
  \BibitemOpen
  \bibfield  {author} {\bibinfo {author} {\bibfnamefont {A.}~\bibnamefont
  {Tebano}} \emph {et~al.},\ }\href@noop {} {\bibfield  {journal} {\bibinfo
  {journal} {Phys. Rev. B}\ }\textbf {\bibinfo {volume} {74}},\ \bibinfo
  {pages} {245116} (\bibinfo {year} {2006})}\BibitemShut {NoStop}%
\bibitem [{\citenamefont {Nanda}\ \emph {et~al.}(2008)\citenamefont {Nanda}
  \emph {et~al.}}]{satpathy-strain-1}%
  \BibitemOpen
  \bibfield  {author} {\bibinfo {author} {\bibfnamefont {B.~R.~K.}\
  \bibnamefont {Nanda}} \emph {et~al.},\ }\href@noop {} {\bibfield  {journal}
  {\bibinfo  {journal} {Phys. Rev. B}\ }\textbf {\bibinfo {volume} {78}},\
  \bibinfo {pages} {054427} (\bibinfo {year} {2008})}\BibitemShut {NoStop}%
\bibitem [{\citenamefont {Nanda}\ \emph {et~al.}(2010)\citenamefont {Nanda}
  \emph {et~al.}}]{satpathy-strain-2}%
  \BibitemOpen
  \bibfield  {author} {\bibinfo {author} {\bibfnamefont {B.~R.~K.}\
  \bibnamefont {Nanda}} \emph {et~al.},\ }\href@noop {} {\bibfield  {journal}
  {\bibinfo  {journal} {Phys. Rev. B}\ }\textbf {\bibinfo {volume} {81}},\
  \bibinfo {pages} {174423} (\bibinfo {year} {2010})}\BibitemShut {NoStop}%
\bibitem [{\citenamefont {Pradhan}\ \emph {et~al.}(2007)\citenamefont {Pradhan}
  \emph {et~al.}}]{pradhan-prl}%
  \BibitemOpen
  \bibfield  {author} {\bibinfo {author} {\bibfnamefont {K.}~\bibnamefont
  {Pradhan}} \emph {et~al.},\ }\href@noop {} {\bibfield  {journal} {\bibinfo
  {journal} {Phys. Rev. Lett.}\ }\textbf {\bibinfo {volume} {99}},\ \bibinfo
  {pages} {147206} (\bibinfo {year} {2007})}\BibitemShut {NoStop}%
\bibitem [{ps-()}]{ps-status}%
  \BibitemOpen
  \href@noop {} {\bibinfo  {journal} {At low temperatures ($T\sim0.005t$)
  strain causes phase separation between F-M and SCO-I or FC-I, as seen in
  experiments~\cite{orb-bias-2, ps-1, ps-2} and theory~\cite{ahn-nat}. However,
  with increasing temperature, such phase separated states rapidly evolve
  either into a global F-M or SCO-I phase and do not affect our finite
  temperature results.}\ }\BibitemShut {NoStop}%
\bibitem [{sca()}]{scaling-status}%
  \BibitemOpen
\bibfield  {journal} {  }\href@noop {} {\bibinfo  {journal} {Using t$\sim
  d^{-2}$ and J$\sim d^{-4}$ known for La$_2$CuO$_4$~\cite{cuprate-scaling},
  only causes quantitative changes. It shifts the metal-insulator transition in
  Fig 2 (a) to 2$\%$ tensile strain and that in Fig 2 (b) to about 2.5$\%$
  compressive strain. These changes are small enough so that our results are
  still easily accesible to experiments}\ }\BibitemShut {NoStop}%
\bibitem [{\citenamefont {Yunoki}\ \emph {et~al.}(2000)\citenamefont {Yunoki}
  \emph {et~al.}}]{yunoki}%
  \BibitemOpen
\bibfield  {journal} {  }\bibfield  {author} {\bibinfo {author} {\bibfnamefont
  {S.}~\bibnamefont {Yunoki}} \emph {et~al.},\ }\href@noop {} {\bibfield
  {journal} {\bibinfo  {journal} {Physical Review Letters}\ }\textbf {\bibinfo
  {volume} {84}},\ \bibinfo {pages} {3714} (\bibinfo {year}
  {2000})}\BibitemShut {NoStop}%
\bibitem [{\citenamefont {Loudon}\ \emph {et~al.}(2002)\citenamefont {Loudon}
  \emph {et~al.}}]{mathur}%
  \BibitemOpen
  \bibfield  {author} {\bibinfo {author} {\bibfnamefont {J.~C.}\ \bibnamefont
  {Loudon}} \emph {et~al.},\ }\href@noop {} {\bibfield  {journal} {\bibinfo
  {journal} {Nature}\ }\textbf {\bibinfo {volume} {420}},\ \bibinfo {pages}
  {797} (\bibinfo {year} {2002})}\BibitemShut {NoStop}%
\bibitem [{\citenamefont {Tokura}(2000)}]{tok-book}%
  \BibitemOpen
  \bibfield  {author} {\bibinfo {author} {\bibfnamefont {Y.}~\bibnamefont
  {Tokura}},\ }\href@noop {} {\bibfield  {journal} {\bibinfo  {journal}
  {\textit{Colossal Magnetoresistive Oxides, Gordon and Breach, Amsterdam}}\ }
  (\bibinfo {year} {2000})}\BibitemShut {NoStop}%
\bibitem [{\citenamefont {Cooper}\ \emph {et~al.}(1990)\citenamefont {Cooper},
  \citenamefont {Thomas}, \citenamefont {Millis}, \citenamefont {Sulewski},
  \citenamefont {Orenstein}, \citenamefont {Rapkine}, \citenamefont {Cheong},\
  and\ \citenamefont {Trevor}}]{cuprate-scaling}%
  \BibitemOpen
  \bibfield  {author} {\bibinfo {author} {\bibfnamefont {S.~L.}\ \bibnamefont
  {Cooper}}, \bibinfo {author} {\bibfnamefont {G.~A.}\ \bibnamefont {Thomas}},
  \bibinfo {author} {\bibfnamefont {A.~J.}\ \bibnamefont {Millis}}, \bibinfo
  {author} {\bibfnamefont {P.~E.}\ \bibnamefont {Sulewski}}, \bibinfo {author}
  {\bibfnamefont {J.}~\bibnamefont {Orenstein}}, \bibinfo {author}
  {\bibfnamefont {D.~H.}\ \bibnamefont {Rapkine}}, \bibinfo {author}
  {\bibfnamefont {S.-W.}\ \bibnamefont {Cheong}}, \ and\ \bibinfo {author}
  {\bibfnamefont {P.~L.}\ \bibnamefont {Trevor}},\ }\href {\doibase
  10.1103/PhysRevB.42.10785} {\bibfield  {journal} {\bibinfo  {journal} {Phys.
  Rev. B}\ }\textbf {\bibinfo {volume} {42}},\ \bibinfo {pages} {10785}
  (\bibinfo {year} {1990})}\BibitemShut {NoStop}%
\end{thebibliography}%


\begin{thebibliography}{10}%
\makeatletter
\providecommand \@ifxundefined [1]{%
 \ifx #1\undefined \expandafter \@firstoftwo
 \else \expandafter \@secondoftwo
\fi
}%
\providecommand \@ifnum [1]{%
 \ifnum #1\expandafter \@firstoftwo
 \else \expandafter \@secondoftwo
\fi
}%
\providecommand \enquote [1]{``#1''}%
\providecommand \bibnamefont  [1]{#1}%
\providecommand \bibfnamefont [1]{#1}%
\providecommand \citenamefont [1]{#1}%
\providecommand\href[0]{\@sanitize\@href}%
\providecommand\@href[1]{\endgroup\@@startlink{#1}\endgroup\@@href}%
\providecommand\@@href[1]{#1\@@endlink}%
\providecommand \@sanitize [0]{\begingroup\catcode`\&12\catcode`\#12\relax}%
\@ifxundefined \pdfoutput {\@firstoftwo}{%
 \@ifnum{\z@=\pdfoutput}{\@firstoftwo}{\@secondoftwo}%
}{%
 \providecommand\@@startlink[1]{\leavevmode\special{html:<a href="#1">}}%
 \providecommand\@@endlink[0]{\special{html:</a>}}%
}{%
 \providecommand\@@startlink[1]{%
  \leavevmode
  \pdfstartlink
   attr{/Border[0 0 1 ]/H/I/C[0 1 1]}%
   user{/Subtype/Link/A<</Type/Action/S/URI/URI(#1)>>}%
  \relax
 }%
 \providecommand\@@endlink[0]{\pdfendlink}%
}%
\providecommand \url  [0]{\begingroup\@sanitize \@url }%
\providecommand \@url [1]{\endgroup\@href {#1}{\urlprefix}}%
\providecommand \urlprefix [0]{URL }%
\providecommand \Eprint[0]{\href }%
\@ifxundefined \urlstyle {%
  \providecommand \doi [1]{doi:\discretionary{}{}{}#1}%
}{%
  \providecommand \doi [0]{doi:\discretionary{}{}{}\begingroup
  \urlstyle{rm}\Url }%
}%
\providecommand \doibase [0]{http://dx.doi.org/}%
\providecommand \Doi[1]{\href{\doibase#1}}%
\providecommand \bibAnnote [3]{%
  \BibitemShut{#1}%
  \begin{quotation}\noindent
    \textsc{Key:}\ #2\\\textsc{Annotation:}\ #3%
  \end{quotation}%
}%
\providecommand \bibAnnoteFile [2]{%
  \IfFileExists{#2}{\bibAnnote {#1} {#2} {\input{#2}}}{}%
}%
\providecommand \typeout [0]{\immediate \write \m@ne }%
\providecommand \selectlanguage [0]{\@gobble}%
\providecommand \bibinfo [0]{\@secondoftwo}%
\providecommand \bibfield [0]{\@secondoftwo}%
\providecommand \translation [1]{[#1]}%
\providecommand \BibitemOpen[0]{}%
\providecommand \bibitemStop [0]{}%
\providecommand \bibitemNoStop [0]{.\EOS\space}%
\providecommand \EOS [0]{\spacefactor3000\relax}%
\providecommand \BibitemShut [1]{\csname bibitem#1\endcsname}%
\bibitem{akahoshi-prl}%
  \BibitemOpen
  \bibfield{author}{%
  \bibinfo {author} {\bibfnamefont{D.}~\bibnamefont{Akahoshi}} \emph{et~al.},\
  }%
  \bibfield{journal}{%
  \bibinfo {journal} {Phys. Rev. Lett.}\ }%
  \textbf{\bibinfo {volume} {90}},\ \bibinfo {pages} {177203} (\bibinfo {year}
  {2003})%
  \bibAnnoteFile{NoStop}{akahoshi-prl}%
\bibitem{pradhan-prl}%
  \BibitemOpen
  \bibfield{author}{%
  \bibinfo {author} {\bibfnamefont{K.}~\bibnamefont{Pradhan}} \emph{et~al.},\
  }%
  \bibfield{journal}{%
  \bibinfo {journal} {Phys. Rev. Lett.}\ }%
  \textbf{\bibinfo {volume} {99}},\ \bibinfo {pages} {147206} (\bibinfo {year}
  {2007})%
  \bibAnnoteFile{NoStop}{pradhan-prl}%
\bibitem{hotta-review}%
  \BibitemOpen
  \bibfield{author}{%
  \bibinfo {author} {\bibfnamefont{E.}~\bibnamefont{Dagotto}} \emph{et~al.},\
  }%
  \bibfield{journal}{%
  \bibinfo {journal} {Physics Reports}\ }%
  \textbf{\bibinfo {volume} {344}},\ \bibinfo {pages} {1} (\bibinfo {year}
  {2001})%
  \bibAnnoteFile{NoStop}{hotta-review}%
\bibitem{harrison}%
  \BibitemOpen
  \bibfield{author}{%
  \bibinfo {author} {\bibfnamefont{W.}~\bibnamefont{Harrison}},\ }%
  \bibfield{journal}{%
  \bibinfo {journal} {\textit{Electronic Structure and the Properties of
  Solids: The Physics of the Chemical Bond, Dover}}}%
   (\bibinfo {year} {1989})%
  \bibAnnoteFile{NoStop}{harrison}%
\bibitem{slater-koster}%
  \BibitemOpen
  \bibfield{author}{%
  \bibinfo {author} {\bibfnamefont{J.~C.}\ \bibnamefont{Slater}}\ and\ \bibinfo
  {author} {\bibfnamefont{G.~F.}\ \bibnamefont{Koster}},\ }%
  \bibfield{journal}{%
  \bibinfo {journal} {Phys. Rev.}\ }%
  \textbf{\bibinfo {volume} {94}},\ \bibinfo {pages} {1498} (\bibinfo {year}
  {1954})%
  \bibAnnoteFile{NoStop}{slater-koster}%
\bibitem{bl-ba-1}%
  \BibitemOpen
  \bibfield{author}{%
  \bibinfo {author} {\bibfnamefont{A.}~\bibnamefont{Machida}} \emph{et~al.},\
  }%
  \bibfield{journal}{%
  \bibinfo {journal} {Phys. Rev. B}\ }%
  \textbf{\bibinfo {volume} {62}},\ \bibinfo {pages} {80} (\bibinfo {year}
  {2000})%
  \bibAnnoteFile{NoStop}{bl-ba-1}%
\bibitem{bl-ba-2}%
  \BibitemOpen
  \bibfield{author}{%
  \bibinfo {author} {\bibfnamefont{O.}~\bibnamefont{Chmaissem}} \emph{et~al.},\
  }%
  \bibfield{journal}{%
  \bibinfo {journal} {Phys. Rev. B}\ }%
  \textbf{\bibinfo {volume} {64}},\ \bibinfo {pages} {134412} (\bibinfo {year}
  {2001})%
  \bibAnnoteFile{NoStop}{bl-ba-2}%
\bibitem{sc-scaling}%
  \BibitemOpen
  \bibfield{author}{%
  \bibinfo {author} {\bibfnamefont{M.~C.}\ \bibnamefont{Aronson}}
  \emph{et~al.},\ }%
  \bibfield{journal}{%
  \bibinfo {journal} {Phys. Rev. B}\ }%
  \textbf{\bibinfo {volume} {44}},\ \bibinfo {pages} {4657} (\bibinfo {year}
  {1991})%
  \bibAnnoteFile{NoStop}{sc-scaling}%
\bibitem{tca}%
  \BibitemOpen
  \bibfield{author}{%
  \bibinfo {author} {\bibfnamefont{S.}~\bibnamefont{Kumar}} \emph{et~al.},\ }%
  \bibfield{journal}{%
  \bibinfo {journal} {Eur. Phys. J.}\ }%
  \textbf{\bibinfo {volume} {B 50}},\ \bibinfo {pages} {571} (\bibinfo {year}
  {2006})%
  \bibAnnoteFile{NoStop}{tca}%
\bibitem{mahan}%
  \BibitemOpen
  \bibfield{author}{%
  \bibinfo {author} {\bibfnamefont{G.~D.}\ \bibnamefont{Mahan}},\ }%
  \bibfield{journal}{%
  \bibinfo {journal} {\textit{ Quantum Many Particle Physics, Plenum Press, New
  York}}}%
   (\bibinfo {year} {1990})%
  \bibAnnoteFile{NoStop}{mahan}%
\bibitem{sk-pm-long-transp}%
  \BibitemOpen
  \bibfield{author}{%
  \bibinfo {author} {\bibfnamefont{S.}~\bibnamefont{Kumar}} \emph{et~al.},\ }%
  \bibfield{journal}{%
  \bibinfo {journal} {Europhys. Lett.}\ }%
  \textbf{\bibinfo {volume} {65}},\ \bibinfo {pages} {75} (\bibinfo {year}
  {2004})%
  \bibAnnoteFile{NoStop}{sk-pm-long-transp}%
\end{thebibliography}%

\end{document}


\title{Supplementary material for ``Theory of strain controlled magnetotransport and stabilization of the ferromagnetic insulating phase in manganite thin "}
\author{Anamitra Mukherjee$^{1, 2}$, William S. Cole$^2$, Patrick Woodward$^3$, Mohit Randeria$^2$ and Nandini Trivedi$^2$ }
\affiliation{$^1$Department of Physics and Astronomy, University of British
Columbia, Vancouver, BC V6T 1Z1, Canada,\\
$^2$Department of Physics, The Ohio State University, Columbus, OH 43210, USA,\\
$^3$Department of Chemistry, The Ohio State University, Columbus, OH 43210, USA}
\maketitle
In this supplementary material we present the following:

\textit{1. Model:} First we give the details of the Hamiltonian and the
parameters used in the main paper in Section I (A) and (B) respectively. We then
discuss the scaling of the hopping and the superexchange parameters under strain
in (C). There we argue why the change in the Mn-O bond length plays a dominant
role in determining these scalings.
 
\textit{2. Method:} We present our method of solution in Section II. The definitions of the spin, charge and orbital structure factors and a brief account of  transport calculations are discussed in Section III. 

\textit{3. Supporting data:} In Section IV we show the magetotransport data (under strain) for a F-M parameter point close to the F-M/SCO-I phase boundary. When contrasted with Fig. 3(a) in the main text, they show that our conclusions regarding the strain response of magnetotransport are independent of the \textit{nature of the insulator-metal boundary} near which the response is calculated.

\section{Hamiltonian and parameters}

\subsection{The manganite Hamiltonian}
\begin{figure}[!b]
\centerline{
\includegraphics[width=18.6cm, clip=true]{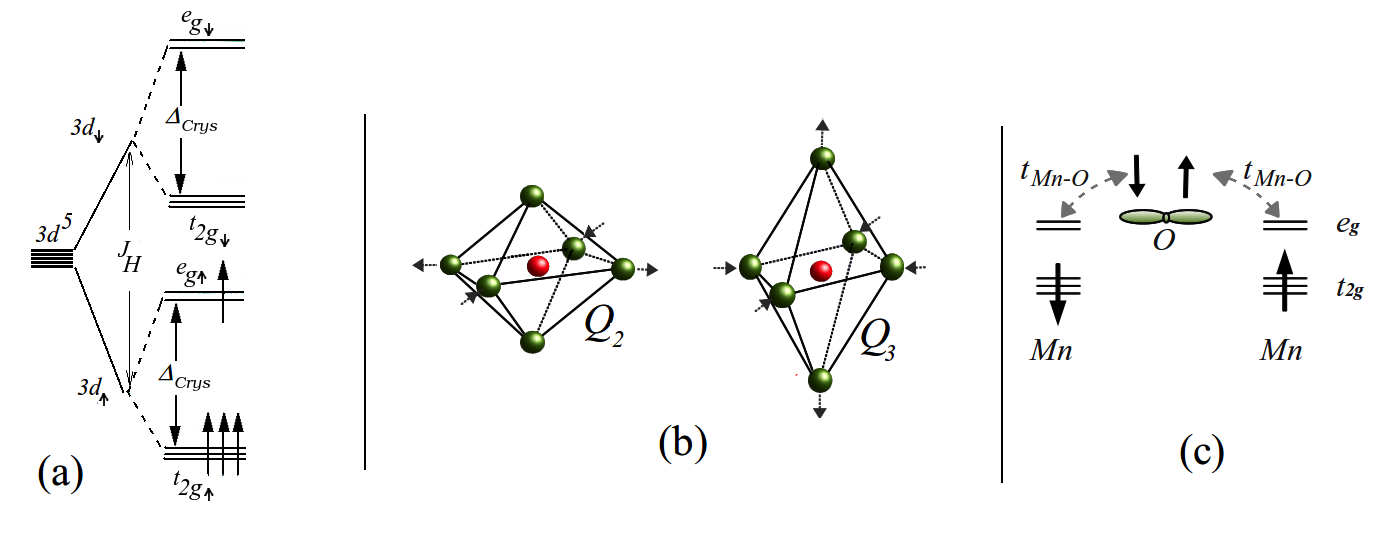}}
\vspace{.2cm}
\caption{{\it Schematics of the relevant interactions incorporated in the 
manganite Hamiltonian:} (a) The manganese level diagram. The largest scale is the Hund's rule scale that splits the up and down manifolds of the Mn $3d$ levels by $J_H$. Next in the energy hierarchy is the cubic crystal
field, that splits that $e_g$ and the $t_{2g}$ states by $\Delta_{Crys}(=10Dq)$. For both the Mn$^{+3}$ and Mn$^{+4}$  states, three electrons occupy the $t_{2g}$ down manifold and are well-localized. The fourth electron, for Mn$^{+3}$, occupies the $e_g$ manifold. (b) As a result of placing a single electron in the degenerate $e_g$ manifold, the system spontaneously undergoes a Jahn-Teller distortion. The electron orbital-pseudospin operator couples to the local distortions $Q_2$, $Q_3$. (c) The effective antiferromagnetic coupling between two nearest neighbor Mn$^{+4}$ atoms. The two $p$ electrons on the oxygen connecting the two Mn sites can gain maximum kinetic energy by hopping to the two Mn sites if their $t_{2g}$  spins are antiparallel. This holds true even when the two Mn atoms are in the $+3$ state, except that the magnitude of this antiferromagnetic superexchange is lowered.
}
\vspace{.2cm}
\label{lm1}
\end{figure}
\textit{ Physical origin of the microscopic interactions: } 

As shown in Fig.~\ref{lm1}(a), the strong Hund's coupling and the crystal field ensure two things: first, that three electrons on the Mn site are well localized in the $t_{2g}$ manifold, forming a $S=3/2$ ``core" spin that may be treated as a classical object, and second, that any electron occupying the $e_g$ level \textit{must} have its spin parallel to the on-site $t_{2g}$ core spin. The electron delocalization happens through the $e_g$ manifold in the metallic state which is modeled by a two-band kinetic energy term with the on-site constraint that the itinerant spin orientation is projected on to the local ``core" spin direction.
This is the double exchange interaction. The spontaneous structural distortions arising from $e_g$ degeneracy-lifting when the $e_g$ manifold is singly occupied, as described in Fig.~\ref{lm1}(b), is incorporated by coupling the $e_g$ electron's orbital pseudospin operator to the Jahn-Teller modes $Q_2$ and $Q_3$. Finally, the Mn-Mn superexchange, due to virtual exchange of the localized oxygen $2p$ electrons with the Mn $e_g$ states, is shown in Fig.~\ref{lm1}(c).\\
\begin{figure}
\centerline{
\includegraphics[width=12.4cm, height=7.6cm, clip=true]{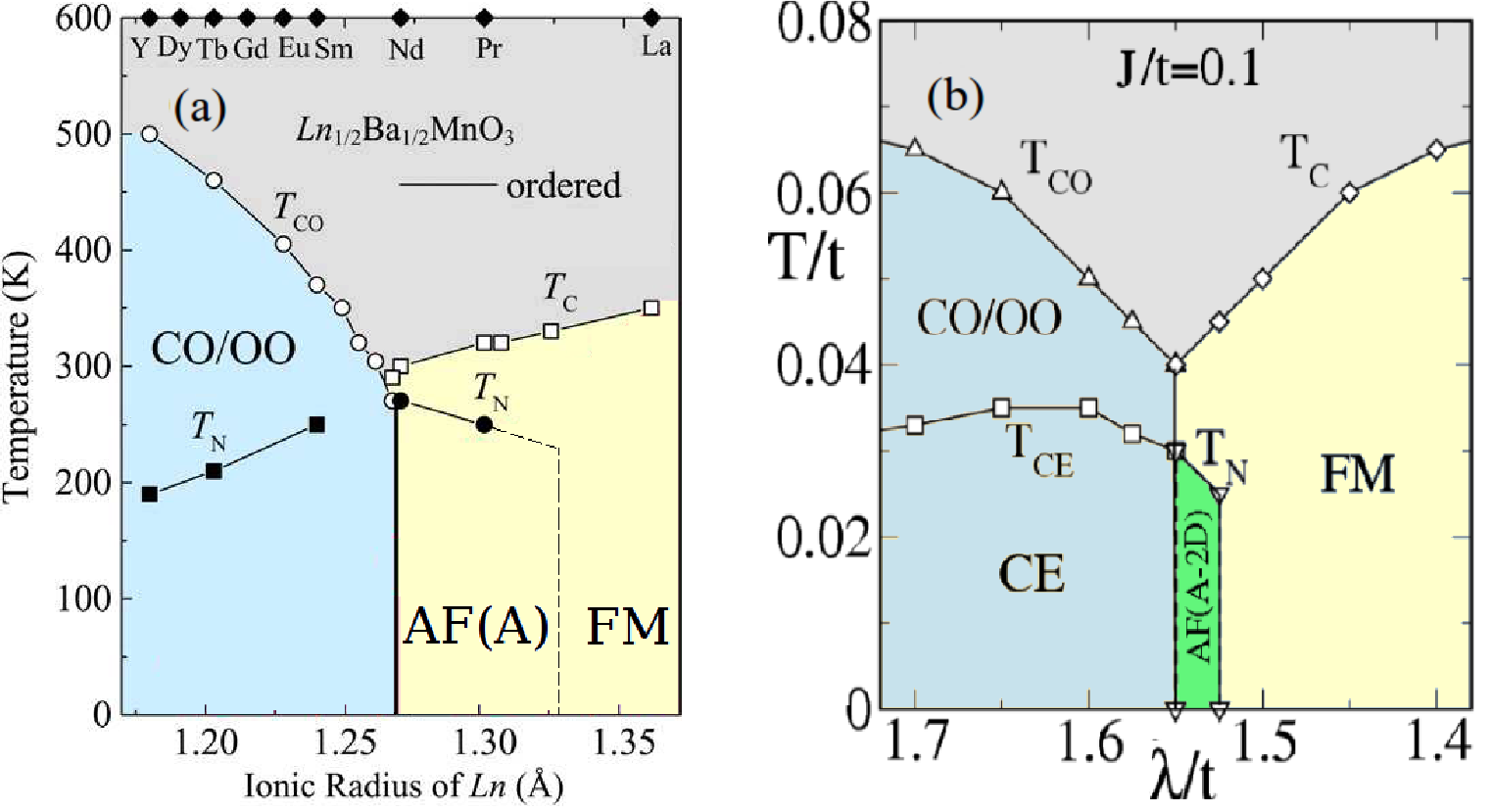}}
\vspace{.2cm}
\caption{ (a) Experimental temperature ($T$) - $r_A$ phase diagram for ordered,
half-doped manganites, from D. Akahoshi \textit{et al.} \cite{akahoshi-prl}. (b)
Previous theoretical work, from K. Pradhan \textit{et al.} \cite{pradhan-prl}:
$T/t - \lambda/t$ phase diagram at $x=0.5$ for ordered half doped manganites.
There is a phase transition between the large $\lambda/t$ or small bandwidth
(BW)  SCO-I and the  small $\lambda/t$ or large BW, F-M on varying $r_A$.
In the experiments, small $r_A$ implies small BW, which corresponds large
$\lambda/t$ in theory. }
\vspace{.2cm}
\label{l0}
\end{figure}
{\it The Hamiltonian} -- Based on the preceding considerations, in this work we consider the microscopic two-band model for $e_g$ electrons with a strong Hund's coupling to $t_{2g}$ core spins in a two-dimensional square lattice. The itinerant electrons are coupled to Jahn-Teller (JT) phonons and the core spins have a nearest-neighbor antiferromagnetic superexchange coupling between them. Explicitly, the Hamiltonian is given by
\begin{eqnarray}
H &=& \sum_{\langle ij \rangle \sigma,\alpha \gamma}
-(t_{\alpha \gamma}^{ij} c^{\dagger}_{i \alpha \sigma} c^{~}_{j \gamma \sigma} +h.c.)
- J_H\sum_i \vec{S}_i \cdot \vec{\sigma}_i 
+ J_{AF}\sum_{\langle ij \rangle} \vec{S}_i \cdot \vec{S}_j \cr
&& +\lambda \sum_i \left( \vec{Q}_i \cdot \vec{\tau}_i - Q_{1i}\rho_i \right)
+ \sum_i \left(\frac{K}{2}\vec{Q}_i^2 + \beta Q_{1i}^2 \right) \cr
&& + \sum_{i,\alpha}\epsilon_{\alpha}n_{i,\alpha} - \mu N - h\sum_i S_{zi}
\end{eqnarray}
where $c$ and $c^{\dagger}$ are the annihilation and creation operators for $e_g$ electrons and $\alpha$, $\beta$ index the two Mn $e_g$ orbitals
($d_{x^2-y^2}$ and $d_{3z^2-r^2}$, labeled $a$ and $b$ throughout). In the unstrained case, $t_{\alpha \beta}^{ij}$ are hopping amplitudes between nearest-neighbor sites with the symmetry-dictated form\cite{hotta-review}:
\begin{align}
&t_{a a}^x = t_{a a}^y \equiv t,		\nonumber \\
&t_{b b}^x = t_{b b}^y \equiv t/3,		\nonumber \\
&t_{a b}^x = t_{b a}^x \equiv -t/\sqrt{3},	\nonumber \\
&t_{a b}^y = t_{b a}^y \equiv t/\sqrt{3}	\nonumber
\end{align}
where $x$ and $y$ refer to the spatial orientation as in Fig. 1 in the main
article. The $e_g$ electron spin operator is given by ${\sigma}^{\mu}_i=\sum_{\sigma
\sigma'}^{\alpha} c^{\dagger}_{i\alpha \sigma} \Gamma^{\mu}_{\sigma\sigma'}
c_{i\alpha \sigma'}$, where the $\Gamma$ are the Pauli matrices. This spin is coupled to the local (classical) $t_{2g}$ spin $\vec{S}_i$ via the Hund's coupling $J_H$, and we assume $J_H/t \gg 1$. Finally, $\lambda$ is the coupling between the JT distortion $\vec{Q}_i = (Q_{2i}, Q_{3i})$ and the orbital pseudospin operator ${\tau}^{\mu}_i = \sum^{\alpha \beta}_{\sigma} c^{\dagger}_{i\alpha \sigma} \Gamma^{\mu}_{\alpha\beta} c_{i\beta \sigma}$. This
same $\lambda$ also sets the scale for the coupling between the breathing mode $Q_1$ and the on site charge density. $K$ is the lattice stiffness. $\beta$, the breathing mode stiffness, is assumed to be sufficiently large that this mode can be neglected. Finally, $h$ is an external magnetic field, assumed to be in the $z$ direction and coupled only to $\vec{S}_i$. We set $t=1$, $K=1$, and treat the $\vec{Q}_i$ and the $\vec{S}_i$ as classical variables which is a a good approximation for 
the manganites\cite{hotta-review}. The chemical potential $\mu$ is adjusted so that the electron density remains at $n=1/2$.

\subsection{Parameter values}

Fig.~\ref{l0} shows the comparison of the experimental, temperature ($T$)-$r_A$ phase diagram \cite{akahoshi-prl} and earlier theoretical work on the $T/t$ -electron-phonon coupling ($\lambda/t$) phase diagram employing the above model \cite{pradhan-prl}. Both the experiment and theory are for ordered, half-doped manganites. The term `ordered' implies alternating layers of $A$=Ba and $A'$=any Lanthanide (Ln), marked on the top of Fig.~\ref{l0}(a). Also the corresponding $r_A$ are shown on the lower axis.

The value of $r_A$ controls the bandwidth (BW) in the experiments. The smaller the $r_A$, the smaller the BW. In the theoretical work, shown in
Fig.~\ref{l0}(b), the change of the BW is incorporated by the ratio $\lambda/t$, where large $\lambda/t$ implies a smaller BW and vice versa. This not only serves as a benchmark of the model and the method of solution, to be described shortly, it also allows us to fix the Hamiltonian parameters by comparing the theoretical and experimental $T_C$ values.

For the electron-phonon and superexchange couplings, we choose the values of $\lambda$ and $J$ to roughly reproduce the thermal magnetic ordering scales seen in experiments. Since we measure all parameters in units of the bandwidth (BW) $t$, the ratios $\lambda/t$ and $J/t$ are what we fix. From Fig.~\ref{l0}, we find that for $\lambda/t\sim1.5$ and $J/t=0.1$, $T_C\sim0.045t$. From this we obtain a $T_C$ of about 225K, which is at least in line with the experimental value near the SCO-I/F-M phase boundary. In doing this conversion we have assumed $t=0.3eV$. In general, there is a consensus that $\lambda/t\sim 1-1.6$ and $J/t\sim 0.05-0.1$ is adequate for reasonably quantitative comparison with experiments for most doping values \cite{hotta-review} and across many families of manganites.

\subsection{Scaling of hopping parameters and superexchange under strain}
(i) Strain puts constraints on the lattice parameter $a$ which in turn changes both the Mn-O bond length $d$ and the Mn-O-Mn bond angle $\phi$ as seen in Fig. 1 (b) in the main text. The $p-d$ overlap integral $V_{pd\sigma}$ scales as $d^{-3.5}$, with $d$ being the center of mass distance between the Mn and O atoms \cite{harrison}. The Mn-O-Mn hopping is therefore proportional to $V^2_{pd\sigma}/\Delta$ and also depends on $\phi$ as $\mbox{cos}^n(\phi)$ (as shown below). Here, $\Delta$ is the energy denominator that depends on the
difference between the Mn and the O states involved. The dependence of the hopping integral on $\phi$ is computed from the Slater-Koster tables\cite{slater-koster} assuming the Mn-O-Mn bond to lie in the
x-y plane.
 
The exponent $n$ is 3 for $t_{aa}$, 2 for $t_{bb}$ and 1 for $t_{ab}$, where
$a=d_{x^2-y^2}$ and $b=3_{z^2-r^2}$. Thus the overall scaling for say, $t_{aa}$ would be $d^{-7}\mbox{cos}^3(\phi)$. Fig. 1 (b) in the main text, shows the relation between the lattice parameter ($a$), the Mn-O bond length ($d$) and the Mn-O-Mn bond angle ($\phi$). Let us assume the strain imposes a change in the lattice parameter by $\delta a$ and this causes a change in $d$ by $\delta d$ and $\phi$ by $\delta \phi$. Thus, $t_{aa} \rightarrow \tilde{t}_{aa} =(d+\delta  d)^{-7}\mbox{cos}^3(\phi+\delta\phi)$. In bulk half doped manganites the value of $\phi$ varies from 160$^\circ$ to 170$^\circ$ in SCO-I, 
Y$_{0.5}$Ba$_{0.5}$MnO$_3$ to F-M, La$_{0.5}$Ba$_{0.5}$MnO$_3$\cite{bl-ba-1,bl-ba-2}. For these values of $\phi$ and assuming small $\delta d/d$ and $\delta \phi$, we can linearize the expression  for $t_{aa}$ in $\delta d/d$ and $\delta \phi$. This gives $\tilde{t}_{aa} \approx t_{aa}\left[1-7\frac{\delta d}{d} - 3\mbox{tan}(\phi)\delta \phi \right]$. The last term controls the extent to which the change in the \textit{unstrained} Mn-O-Mn bond angle will effect the hybridization. For materials with $\phi\sim{160^\circ}$, the ratio $| 3 \mbox{tan}(\phi)\cdot d \frac{\delta \phi}{7\delta d} |$ is $|0.156 \cdot d\frac{\delta\phi}{\delta d}|$. To estimate this ratio we need typical data for perovskite structure under strain. Unfortunately, to the best of our knowledge, such data does not exist for the manganites. So as an estimate we assume a reasonable $|\delta d/d| \sim 0.01$ at 1$\%$  strain. Then it can be shown that for $\delta \phi<4^\circ$, the
in-plane hopping increases (decreases) with $\delta d$ being negative
(positive), while the bond angle dependence causes only quantitative change. Thus we neglect the $\delta \phi$ dependence in our calculation for small strain.

In Fig. 1 (b) in the main text, $a=2d\sin(\phi/2)$.  However in the spirit of
above approximation we simply consider, $a=2d$, or, $2\delta d=\delta a$. So
for the x-y plane, e$_{\parallel}$ = $\delta d/d$. Thus the hopping in the (x-y)
plane scales with strain as $t_{\parallel}\sim
t\left(1-7e_{\parallel}\right)$, while the out of plane hopping scales as
$t_{\bot}\sim t\left(1-7e_{\bot}\right)$.

(ii) The scaling of the superexchange, $J$, with strain depends of the scaling
of the \emph{fourth} power of $V_{pd\sigma}$ under interatomic separation. Using
considerations similar to above it is easy to show that,
$J_{\parallel}\rightarrow \tilde{J_{\parallel}}\sim (J_{\parallel})(1-14e_{\parallel}))$ with strain. Similar scaling has been
observed on application of pressure in La$_2$CuO$_4$\cite{sc-scaling}. The anisotropy in the hopping is also reflected in anisotropic antiferromagnetic scales $J_{\parallel}$ and $J_{\bot}$.
\begin{figure}
\centerline{
\includegraphics[width=14.6cm, height=6.2cm, clip=true]{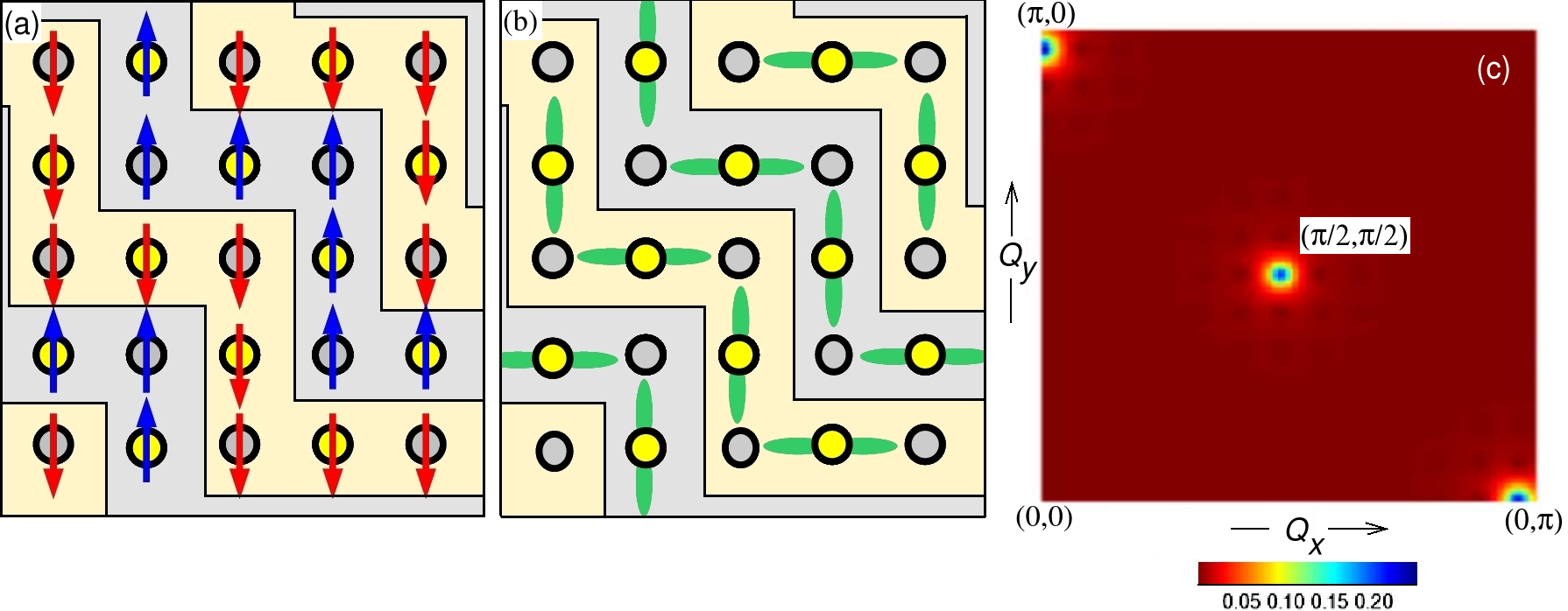}}
\vspace{.2cm}
\caption{The schematic of the SCO-I phase: (a) shows the planar checkerboard charge order (CO) with larger charge on the 'yellow' sites and the CE type spin order (zig-zag ferromagnetic chains coupled antiferromagnetically). (b) shows the alternating $d_{x^2-r^2}/d_{y^2-r^2}$ orbital order (on the sites with larger charge density). (c) The color map of the magnetic structure factor, $S(\vec{q})$ in the momentum space of our numerical data for the CE phase at low T. The $\vec{q}=(0, \pi),(\pi,0),$ and $(\pi/2,\pi/2)$ peaks are clearly seen.
}
\vspace{.2cm}
\label{l4}
\end{figure}
\section{Method of solution}

As mentioned above, we treat the core spins classically and the phonons in the
adiabatic limit. Thus the solution of the Hamiltonian amounts to determining the
configuration of the classical core spin and phonon background (for a fixed
fermion density and temperature) that minimizes the free energy. To do this, we
perform an exact diagonalization (ED) of the itinerant electron system for each
configuration of the background spins and phonons. The Boltzmann weight of the
combined electron, spin, and phonon system is calculated, and new spin/phonon
configurations are sampled through classical Monte Carlo (MC). At a fixed
temperature a Monte-Carlo system sweep consists of visiting every site of the
system once in a sequential manner and performing the above mentioned update.

The combined algorithm (ED+MC) is numerically rather costly, since the exact
diagonalization must be performed at every step and the cost scales as
$\mathcal{O}(N^3)$ with $N$ the number of lattice sites. Additionally, with a
sequential system sweep, the cost of a  Monte-Carlo system sweep scales as $N^4$
at each temperature.  To reach reasonably large system sizes, we employ a
recently developed variation \cite{tca} of real space ED+MC to that allows a
linear scaling with the system size. This technique defines a region (cluster)
around the site at which one attempts a MC update and accepts or rejects the
update based on the energy change only in the cluster rather than diagonalizing
the full system. For the calculation of observables, we diagonalize the full
system, after the system has equilibrated using the above algorithm (known as
the `traveling cluster approximation' (TCA)). This adds only a few hundred full
system diagonalizations to the computation cost.

Within TCA, at each temperature, the computation cost of ED for a system with
$N$ sites is  $\mathcal{O}(N_{C}^{3})$, where $N_{C}$ is the fixed cluster
size. Thus the cost of a full sweep of the lattice is $NN^3_c$ or {\it linear}
in $N$ as opposed to $N^{4}$. Using this technique we have accessed sizes up to
$24^{2}$ as opposed to the practical limit of $\sim 8^{2}$ within conventional
ED+MC in 2 dimensions. In the present work on two dimensional systems, we employ
a $8^2$ travelling cluster on system sizes between $16^2$ (for most data
presented here) to $24^2$. In studies such as this, where
real space phase separation may play a vital role in determining transport
responses, it is crucial to have access to system sizes that are large enough
to capture coexisting phases on the lattice.

We perform 4000 Monte-Carlo system sweeps at every temperature. Of these the
first 2000 are used to equilibrate the spin and phonon variables. 
Every 100$^{th}$ step from the remaining 2000 is used to calculate various
quantities of interest.

For calculating the response to an applied magnetic field, we repeat the above
annealing process in the presence of the magnetic field. For example to
calculate magnetoresistance at a given value of strain, we compare the
resistivity calculated with and without a magnetic field.

\section{Characterization of ordered phases $\&$ transport}

\subsection{Structure factors}
To characterize the phases as a function of strain, temperature, and magnetic
field, we employ a number of static structure factors. The magnetic structure
factor, defined by $S(\vec{q})=\sum_{ij} \langle \vec{S}_i \cdot \vec{S}_j
\rangle e^{i \vec{q} \cdot (\vec{r}_i-\vec{r}_j)}$, has a sharp peak at
$\vec{q}=(0,0)$ for the ferromagnetic state and peaks at $\vec{q}=(0, \pi),(\pi,
0),$ and $(\pi/2,\pi/2)$ for the in-plane `CE' type spin order. The long range
charge order is characterized by the structure factor $D_Q(\vec{q})=\sum_{ij}
\langle \vec{Q}_i \cdot \vec{Q}_j \rangle e^{i \vec{q}.(\vec{r}_i-\vec{r}_j)}$.
The in-plane checkerboard charge order is indicated by a peak at
$\vec{q}=(\pi,\pi)$. The schematic of the SCO-I phase in shown on Fig.~\ref{l4}.
(a) shows the a-b plane checkerboard charge order and the antiferromagnetically
coupled zig-zag ferromagnetic chains. (b) shows the schematic of the alternating
orbital order. (c) shows the color map of the magnetic structure factor,
$S(\vec{q})$ in the momentum space of our numerical data for the SCO-I order at
low T. The $\vec{q}=(0, \pi),(\pi,0),$ and $(\pi/2,\pi/2)$ peaks are clearly
seen signifying the stabilization of the CE
phase.

\begin{figure}
\centerline{
\includegraphics[width=12.6cm, height=7.2cm, clip=true]{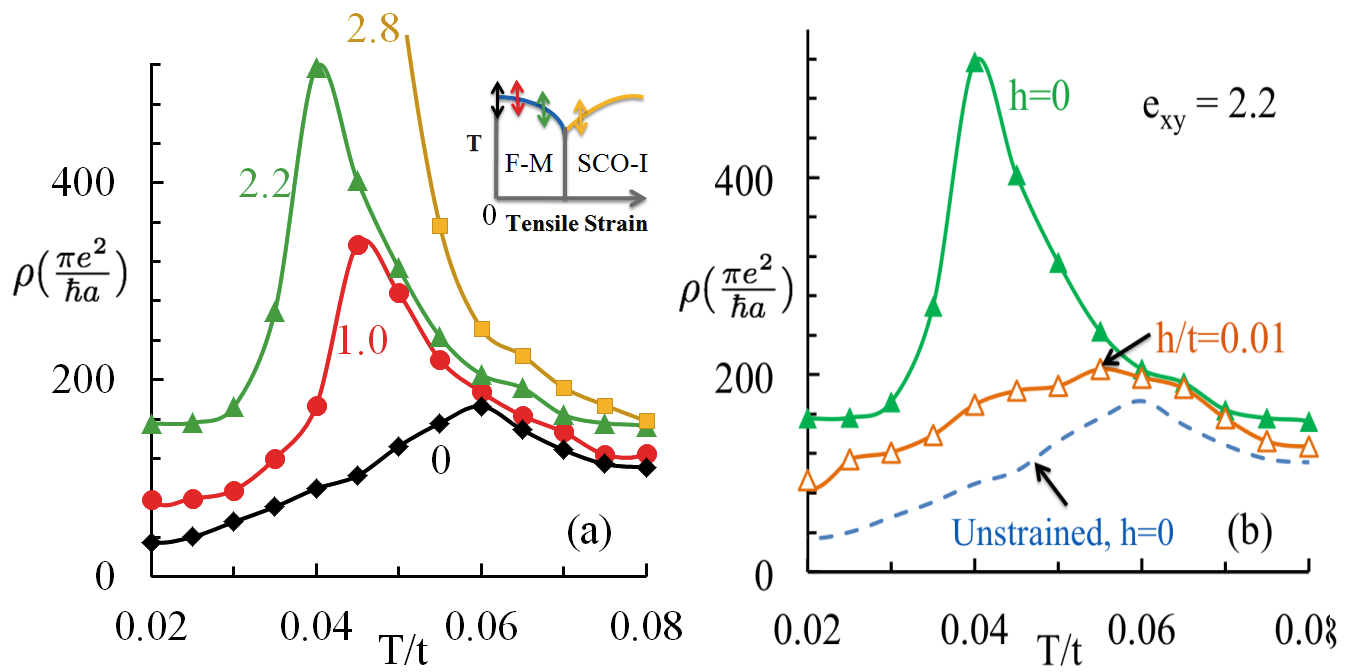}}
\vspace{.2cm}
\caption{The resistivity vs temperature (a) and the response of the resistivity
to magnetic fields (b). The data is shown for a F-M which is close to a tensile
strain driven F-M/SCO-I transition. In the inset in (a), the location of
T$_{CO}$ (yellow) and T$_c$ (rest) for the different values of strain, as
indicated in main panel, are shown in the same color by the double headed
arrows. We find that the strain and field response of the dc resistivity is
qualitatively similar to the results for a F-M close to the tensile strain
induced F-M/FC-I transition in the main text. Also shown in (b) for comparison,
is the unstrained resistivity vs temperature at zero field (dashed line).}
\vspace{.2cm}
\label{l5}
\end{figure}

\subsection{Transport calculations}
The d.c conductivity $\sigma_{dc}$ is estimated by the Kubo-Greenwood
expression \cite{mahan} for the optical conductivity. In a non-interacting
system:
\begin{equation}
\sigma(\omega)=\frac{\pi e^2}{N\hbar a}
\sum_{\alpha,\beta} (n_{\alpha} - n_{\beta})
\frac{|f_{\alpha \beta}|^2}{ \epsilon_{\beta} - \epsilon_{\alpha} }
\delta(\omega - (\epsilon_{\beta} - \epsilon_{\alpha}))
\end{equation}
The $f_{\alpha\beta}$ are the matrix elements of the current operator,
e.g., $\langle \psi_{\alpha} | j_x | \psi_{\beta} \rangle$, and the current
operator itself (in the tight-binding model) is given by $j_x = i t a e
\sum_{i, \sigma} (c^{\dagger}_{{i + a\hat{x}},\sigma} c_{i, \sigma} - h.c)$.
The $\psi_{\alpha}$ are single-particle eigenstates, and $\epsilon_{\alpha}$ are
the corresponding eigenvalues. The $n_{\alpha}=f(\mu - \epsilon_{\alpha})$ are
Fermi factors. $a$ in the lattice spacing.

We can compute the low-frequency average, $\sigma_{av}(\mu, \Delta \omega, N) =
(\Delta \omega)^{-1} \int_0^{\Delta \omega} \sigma(\mu, \omega, N) d\omega$,
using periodic boundary condition in all directions. The averaging interval is
reduced with increasing $N$, with $\Delta \omega \sim B/N$. The d.c.
conductivity is finally obtained as $\sigma_{dc}(\mu) = {\lim}_{L \rightarrow
\infty} { \sigma}_{av}(\mu, B/L, L)$. The chemical potential is set to target
the required electron density $n$. This approach to d.c. transport calculations
has been benchmarked in a previous work\cite{sk-pm-long-transp}.

\section{Tensile strain induced metal to insulator transition across the
F-M/SCO-I boundary}
Here we show that qualitatively our conclusions are independent of the choice
of the metal-insulator boundary near which we calculate  magnetotransport. For
this we present the magnetotransport for a F-M with ($\lambda/t=1.5, J/t=0.1$)
which is close to tensile strain driven F-M/SCO-I boundary. From
Fig.~\ref{l5}(a) we can easily deduce that the enhancement in $\%$ MR
with tensile strain is qualitatively similar to that seen in Fig. 3(a) in the
main text. Also shown in Fig.~\ref{l5}(b) is the magnetic field induced colossal
suppression of resistivity for 2.2$\%$ tensile strain. Fig. 3(c) in the main
text, are constucted from similar data.

\bibliography{mnbib}